\documentclass{webofc}
\usepackage[varg]{txfonts}   
\usepackage{float,amsmath,amssymb}
\begin{document}
\title{Theory progress at Strange Quark Matter 2021}

%
%

\author{\firstname{Björn} \lastname{Schenke}\inst{1}\fnsep\thanks{\email{bschenke@bnl.gov}}
}

\institute{Physics Department, Brookhaven National Laboratory, Upton, NY 11973, USA}

\abstract{%
 I review recent theory progress reported at the 19th International Conference on Strangeness in Quark Matter (SQM), and discuss open questions to be addressed by the coming editions of SQM.
}
\maketitle
\vspace{-0.5cm}
\section{Introduction}
\label{intro}
The focus of the "Strange Quark Matter" (SQM) conference series is the understanding of strangeness and heavier flavor production and evolution in strongly interacting matter, as generated in heavy ion collisions or present in neutron stars. The conference also addresses related questions, such as that of the detailed structure of the phase diagram of quantum chromo dynamic (QCD) matter, including the potential existence of a critical point. Furthermore, the process of hadronization, the conversion of free quarks and gluons into color neutral hadrons, as well as the determination of QCD transport coefficients were major topics discussed at this year's conference. In this proceeding I will focus on new theory developments and important open questions that remain for future SQM conferences.

\section{Equation of state and phase diagram}
\label{sec-eos}
The QCD equation of state is well determined from lattice QCD for zero baryon chemical potential $\mu_B$. For finite baryon chemical potential, however, the sign problem prevents direct evaluation using standard Monte Carlo methods. Different approaches to reach increasingly large $\mu_B$ are being followed, for example that of Taylor expansion around $\mu_B=0$. This method suffers from slow convergence and bad signal to noise ratio for the higher order expansion coefficients. Recent work was presented that reorganizes the expansion scheme \cite{Borsanyi:2021sxv} and can determine thermodynamic observables up to $\mu_B=3.5 T$. One should note that singularities in the complex $\mu_B$ plane will make the expansions fail, and there are various indications that this happens at $\mu_B/T\gtrsim 1.5$ \cite{Mondal:2021jxk}. Understanding the nature of the singularity responsible for such a breakdown is a topic of great interest. It will need detailed quantitative studies involving careful finite-volume scaling analyses.

Besides lattice QCD, model calculations also allow to explore the equation of state at finite chemical potentials. Results from a chiral mean field model were presented \cite{Motornenko:2020yme}, and it was found that results for susceptibilities were dependent on the repulsive interactions between hadrons, in particular the hard core repulsion of hyperons. Comparison to lattice data where available, also lead to the conclusion that the excluded volume for different particles matters \cite{Goswami:2020yez}, in particular that strange baryons have smaller excluded volumes than non-strange ones \cite{Motornenko:2020yme}. 

The presence of strange baryons, or hyperons, affect the equation of state, and in turn can affect neutron star properties. In fact, the ``hyperon puzzle'' states that when including hyperons, the mass-radius relation of neutron stars becomes incompatible with experimental observations. There are many suggestions for a way out, including modified hyperon interactions, effects of meson condensates or quark matter, or even dark matter effects or modified gravitational theories \cite{Tolos:2020aln}. New machine learning techniques were presented as another way to constrain the neutron star equation of state \cite{Fujimoto:2021zas}. From this method, the presence of a first order transition could not be excluded. Deep learning was also trained to determine the phase transition order in a linear sigma model \cite{Jiang:2021gsw}, which in the future could lead to new methods to apply to experimental data. 

The search for the QCD critical point relies heavily on observations of fluctuations of conserved charges, for which extrema are expected if the system passes close to the critical point. Model improvements to allow for a better comparison of fluctuation observables with experiment were presented \cite{Vovchenko:2021kxx}. This included hydrodynamics together with an excluded volume hadron resonance gas model, matched to lattice QCD susceptibilities. Proton cumulants in the experimental acceptance were first calculated in the grand-canonical limit, then corrections for exact baryon number conservation were applied. This method allows for example a direct comparison of net-proton cumulants, which turn out to be significantly different from net-baryon cumulants. It remains an open question whether the experimentally observed extrema in these observables as a function of beam energy indicate any criticality. Effects of non-equilibrium (in the initial state \cite{Dore:2020jye} and the evolution \cite{Kitazawa:2020kvc}) on fluctuation observables were also addressed, and can be of great significance. 

New lattice QCD calculations of electric charge and strangeness cumulant ratios were presented \cite{Bollweg:2020pjb}. They allow the extraction of freeze-out chemical potentials by comparison to experimental data.
Freeze-out was a concern in several more presentations, all of which agreed that strange particles freeze out at a higher temperature than non-strange ones \cite{Alba:2020jir, Inghirami:2021zja, Chen:2020zuw}.

\section{Transport coefficients and medium effects}
\label{sec-transport}

Besides the diffusion coefficients for strange and heavy quarks, which we will focus on below, the properties of the bulk medium, characterized by the shear and bulk viscosity to entropy density (or enthalpy) ratios were also addressed. In particular the temperature and chemical potential dependence of these quantities is a matter of great current interest. A minimum of the effective $(\eta/s)(\sqrt{s})$ was found by analysis of the RHIC beam energy scan data using a scaling function \cite{lacey}, the shear viscosity to enthalpy ratio was computed in a hadron resonance gas model \cite{McLaughlin:2021dph} as a function of $T$ and $\mu_B$, and effects of trajectories in the phase diagram were studied. Also, $(\eta/s)(T)$, $(\zeta/s)(T)$, and the electrical conductivity were computed within a quasiparticle model \cite{Mykhaylova:2020pfk}.

Lattice QCD provides a powerful tool to compute heavy quark diffusion coefficients. Both spatial and momentum diffusion coefficients for charm and bottom quarks were determined using both hadronic and gluonic correlators \cite{Ding:2018uhl,Altenkort:2020fgs}. 
Alternatively, one can use effective theories to determine diffusion coefficients for heavy flavors, such as those based on chiral and heavy-quark spin-flavor symmetries within the imaginary-time formalism \cite{Montana:2020lfi}. Including thermal and off-shell effects, it is found within such model that the D-meson spatial diffusion coefficient matches smoothly to the latest results of lattice-QCD calculations \cite{Torres-Rincon:2021yga}.

All transport coefficients are important inputs for calculations of the bulk medium evolution and in-medium evolution of heavy-quarks. There was some focus on transport theory with results presented for comparisons of microscopic transport simulations and hydrodynamics. This study indicates that charm quarks only approach local thermal equilibrium at small transverse momentum, even though they acquire an elliptic flow that is comparable to light-quark hadrons also at larger $p_T$ \cite{Ding:2021ajz}.
Transport simulations were also performed at fixed $\eta/s$, which can be compared to hydrodynamics, pushed far off equilibrium, and can incorporate different hadronization prescriptions, to compare e.g. microscopic implementations to Cooper-Frye \cite{galesi}. Extensions to the same transport descriptions to include vorticity and electromagnetic fields were also presented \cite{plumari}. Such solutions to the Boltzmann equation were also used to describe both the evolution of the medium and heavy quarks within, using fluctuating initial conditions. Diffusion coefficients for charm and bottom quarks consistent with lattice QCD results, allow to describe the experimental data for $R_{AA}$ and $v_2$ for example \cite{sambataro}.

Antikaon properties in nuclear matter were also computed using a unitarized coupled-channel model in dense and hot matter (or G-matrix approach) based on the SU(3) meson-baryon chiral Lagrangian. This can be implemented in the parton hadron string dynamics (PHSD) transport code and observables, including momentum anisotropies, calculated \cite{Song:2020clw}. These calculations show, by comparison to data from the KaoS, FOPI and HADES Collaborations, that medium modifications of (anti)kaon properties are necessary to explain the data. Also effects of the strong magnetic field on heavy quark transport were investigated, yielding mainly a directional anisotropy \cite{kurian}. 

The behavior of quarkonium in the medium was studied by solving a Lindblad equation for the evolution of the heavy-quarkonium reduced density matrix, derived using potential non-relativistic QCD and the formalism of open quantum systems \cite{delorme, Brambilla:2021wkt}. In combination with hydrodynamic simulations for the medium, this calculation leads to results for the suppression and elliptic flow of the $\Upsilon(1S)$, $\Upsilon(2S)$, and $\Upsilon(3S)$, as functions of centrality and transverse momentum, in very good agreement with the experimental data \cite{Brambilla:2021wkt}.

Medium modifications of bottomonium spectral functions, namely the mass shift and broadening, were also computed on the lattice \cite{Larsen:2019zqv}. While masses are barely modified, the widths of the $\Upsilon(1S)$, $\Upsilon(2S)$, and $\Upsilon(3S)$ increase significantly with temperature. 
These lattice results were also used as input in a deep learning analysis to extract the heavy quark potential in the quark gluon plasma \cite{Shi:2021qri}, and the imaginary part turned out to be quite different from that of the hard thermal loop potential used in \cite{Brambilla:2021wkt}. As for heavy quarks, also for charmonium states effects of strong electromagnetic fields, as well as vorticity fields, were considered and studied via the two-body Schroedinger equation \cite{Chen:2020xsr}. Both mass and shape of the charmonium were significantly affected by the fields.

There has been a continued difficulty to simultaneously describe the medium suppression, quantified by $R_{AA}$ and the elliptic anisotropy of high momentum probes, both light and heavy. At this conference, the Dynamical Radiative and Elastic Energy Loss Approach (DREENA) was used with both Bjorken expansion and a 3D hydrodynamic medium to simultaneously describe the two observables in heavy ion collisions at LHC \cite{Zigic:2018ovr,Zigic:2019sth}. The authors give the proper description of the parton-medium interaction as the main ingredient of the framework that is responsible for its success. Somewhat in contrast, in previous studies the importance of fluctuations for the resolution of the $R_{AA}-v_2$ puzzle was highlighted \cite{Noronha-Hostler:2016eow}. It remains to be seen what the main ingredient is and what the DREENA framework will give when fluctuations are included.

\section{Hadronization}
\label{sec-hadron}
One interesting experimental observation is strangeness enhancement in high multiplicity systems as measured by the ALICE Collaboration \cite{ALICE:2016fzo,ALICE:2019avo}. Different theoretical models with very different underlying physics pictures are able to describe the experimental data. This includes PYTHIA with ``rope hadronization'' (at least for p+p collisions) \cite{Nayak:2018xip}, a (hydrodynamic) core+corona model \cite{Kanakubo:2019ogh}, and thermal (statistical hadronization) model with canonical suppression included \cite{ALICE:2018pal,Wheaton:2004qb}. It remains an open question as to which picture reflects nature most accurately, in particular for the smaller systems considered. 

New results from the above mentioned statistical hadronization model, including an extension to include charm degrees of freedom were also presented \cite{Andronic:2021erx}. It was shown that the measured multiplicities of single charm hadrons in heavy ion collisions at LHC energies can be well described with the same thermal parameters as for the light quark flavor hadrons. Production probabilities for doubly- and triply-charmed hadrons are predicted to exhibit a characteristic  enhancement hierarchy. The model also allows to determine transverse momentum spectra by combining it with blast wave models \cite{Andronic:2021erx,Harabasz:2020sei}.

An interesting observation was presented that indicates a non-universality of hadronization when comparing $e^+(p)+e^-$ collisions with p+p collisions. It appears that charm quarks are ``redistributed'' from mesons to baryons as one goes from collisions involving electrons to p+p collisions \cite{ALICE:2021dhb}. Again, a variety of theoretical models, ranging from PYTHIA with ``enhanced color reconnections'', to a statistical hadronization model with additional excited baryon states, to a combined coalescence and fragmentation model, can describe this enhancement of the charmed baryon yield \cite{ALICE:2020wfu}. The latter model also does best in describing the heavier baryon states, such as $\Xi_c^0$ and $\Xi_c^+$.

After hadronization, hadrons and resonance states interact, decay, and recombine. This is often described using so called hadronic afterburners. Their effects were studied in detail in \cite{Oliinychenko:2021enj}. The afterburner leads to a suppression of most resonances (enhancement of some), an increase in their mean transverse momentum, and a suppression of their $v_2$ at small $p_T$. Similar results were also found by comparing EPOS with EPOS+UrQMD \cite{Kiselev:2020ktr,Knospe:2015nva}. 

Finally the production of light nuclei was addressed within a three-fluid dynamic model with coalescence and UrQMD \cite{Kozhevnikova:2020bdb}, as well as within the Parton-Hadron-Quantum-Molecular Dynamics model \cite{Glassel:2021rod}. Within the latter it was found that clusters are formed shortly after elastic and inelastic collisions have ceased, and behind the front of the expanding energetic hadrons, which allows them to survive.

\section{Polarization and spin}
\label{sec-spin}

The large global angular momentum (of e.g. $L=10^7 \hbar$ in $2.76\,{\rm TeV}$ collisions at LHC for impact parameter $b=7\,{\rm fm}$) in heavy ion collisions generates an enormous vorticity of the order of $10^{21}\,{\rm s}^{-1}$. Spin orbit coupling leads to polarization of produced particles. For the hyperon, the decay products of its weak decay carry information on its polarization in their angular distribution. The phase space averaged polarization measured in this way at RHIC is well described by models. Calculations using AMPT were presented and able to describe the global polarization of $\Lambda$, $\Xi^-$, and $\Omega^-$ hyperons for $200\,{\rm GeV}$ collisions at RHIC \cite{Li:2021zwq}. Predictions for lower energies were also made. The ordering $P_{\Omega^-}>P_{\Xi^-}>P_{\Lambda}$ was found, where $P_X$ is the polarization of the respective particle $X$.

It is puzzling that the angle differential polarization has the opposite sign in experiment compared to calculations. An attempt to remedy this situation by including a previously neglected shear induced polarization was presented \cite{Fu:2021pok,Liu:2021uhn}. It leads to an improvement only if in addition the $\Lambda$ inherits the strange quark's polarization and that spin polarization is frozen from the hadronization on. Further developments concerning the spin polarization were presented, including new proposals on how best to measure the hyperon polarization, specifically under consideration of the different rest frames involved \cite{Florkowski:2021pkp}. Predictions for the collision energy dependence of $\Lambda$ polarization were also presented, indicating a maximum at $\sqrt{s}\approx 7.7\,{\rm GeV}$ \cite{Guo:2021uqc}. Also the effect of vorticity on the spin alignment of vector mesons was addressed, in particular using a quark coalescence model \cite{Sheng:2020ghv,Sheng:2019kmk}. Effective vector meson fields and vorticity fields turn out to be important for the spin alignment in this model. There remain many open questions concerning spin in heavy ion collisions. What effect does rotation have on strongly interacting matter (equation of state, phase structure, vector mesons)? Can we develop a complete spin transport theory or spin hydrodynamics to understand all aspects of spin in heavy ion collisions? The field will be working towards answering these and other questions discussed in this short overview by the next SQM. 

\section{Conclusions}
Besides the tremendous experimental advances \cite{bellini}, the theoretical progress towards understanding of strange and heavy flavor production reported at this edition of SQM has been significant. As discussed in this brief summary, many open questions remain and new ones are emerging as we learn more both from experiment and new theoretical developments.

\section*{Acknowledgments}
B.P.S. is supported under DOE Contract No. DE-SC0012704. 

\vspace{-0.2cm}
\bibliography{spires}

\end{document}